\newif\ifsubmission
\submissionfalse

\ifsubmission
  \documentclass{submission}
\else
  \documentclass[11pt]{article}
  \usepackage{fullpage}
  \usepackage{theorem}
\fi
\usepackage{amsmath,amssymb,mathrsfs}
\usepackage{graphicx}
\usepackage{subcaption}
\usepackage{hyperref}
\usepackage{xcolor}
\usepackage[plain]{algorithm2e}

\newcommand{\db}{\mathbf{x}}
\newcommand{\att}{x}
\newcommand{\id}{i}
\newcommand{\ID}{\mathcal{I}}
\newcommand{\q}{q}
\newcommand{\err}{e}
\newcommand{\Err}{\mathcal{E}}
\newcommand{\zo}{\{0,1\}}
\newcommand{\ans}{a}

\newcommand{\loans}{\mathsf{loans}}
\newcommand{\banking}{\mathsf{banking}}
\newcommand{\loanstatus}{\mathsf{loanStatus}}
\newcommand{\clientid}{\mathsf{clientId}}
\newcommand{\clientids}{\mathsf{clientIds}}

\begin{document}

\ifsubmission
\else
  \title{Linear Program Reconstruction in Practice}
  \author{Aloni Cohen\thanks{MIT, \texttt{aloni@mit.edu}. Work partially done when the author was visiting Georgetown University. Supported by NSF Graduate Research Fellowship, Facebook Fellowship, NSF Project CNS-1413920. Code and data used in this work are available at \url{https://github.com/a785236/Linear-Program-Reconstruction}.}
  \and
  Kobbi Nissim\thanks{Department of Computer Science, Georgetown University, \texttt{kobbi.nissim@georgetown.edu}.}}
  \date{\today}
\fi

\ifsubmission
\else
  \maketitle
  \begin{abstract}
    We briefly report on a successful linear program reconstruction attack performed on a production statistical queries system and using a real dataset. The attack was deployed in test environment in the course of the Aircloak Challenge bug bounty program and is based on reconstruction algorithm of \cite{DMT07}. We empirically evaluate the effectiveness of the \cite{DMT07} algorithm and the related \cite{DiNi} algorithm with various dataset sizes, error rates, and numbers of queries in a Gaussian noise setting.
  \end{abstract}
\fi

\section*{Introduction}
Responding to public and legislation pressures, companies are seeking practical and usable technological solutions to their data privacy problems. Larger corporations often employ Chief Privacy Officers and teams of privacy engineers to develop custom data privacy solutions. Some of these companies, including Google, Apple, and Uber, are recently experimenting with provable approaches to privacy using cryptography and differential privacy, which---at their current stage of development---require significant research and engineering efforts.

But not all companies have the means and technological sophistication to adopt this sort of bespoke approach to privacy. This void is being filled by a growing industry of companies selling off-the-shelf data privacy solutions, many of which aim to \emph{anonymize} or \emph{de-identify} sensitive data.
These companies often advertise their anonymization products as not only preventing the disclosure of sensitive data and but also ensuring compliance with relevant privacy laws, including HIPAA, FERPA, and GDPR.\footnote{HIPAA is the US Health Insurance Portability and Accountability Act. FERPA is the US Family Educational Rights and Privacy Act. The GDPR is the EU General Data Protection Regulation.}
Lack of transparency surrounds some of these technologies. Even when disclosed, the technical underpinnings of many privacy protection claims are heuristic and hence hard to evaluate.

Heuristic approaches to data privacy are not typically ruled out by data privacy regulations.
Furthermore, these regulations are not typically interpreted to require a strong level of protection from data privacy technology.
For example, the EU's General Data Protection Regulation limits its scope to those ``means reasonably likely to be used'' to re-identify data~\cite{gdpr}. A report from the UK's Information Commissioner's Office interprets similar language in earlier legislation as requiring protection against ``motivated intruders''; alas, these intruders are assumed to lack both ``any prior knowledge'' and ``specialist expertise''~\cite{ico}. This policy approach allows practitioners to argue that data privacy technology can be deployed even when they can be theoretically demonstrated to be vulnerable to attacks, as \emph{purely theoretical} attacks plausibly fall outside the scope of the relevant regulations

We believe that this is an unhealthy state of affairs. However, one can hope to affect the legal interpretation of existing regulations by implementing theoretical attacks and demonstrating their practicality. As a striking example, the decision to use differential privacy for the 2020 Decennial Census in the US was largely motivated by the Census Bureau's realization that traditional statistical disclosure limitation techniques may be vulnerable to practical reconstruction attacks~\cite{death_knell}.

\ifsubmission
\else
\paragraph{Reasoning about privacy.}
\fi Academics have developed paradigms to reason formally about certain aspects of data privacy. Among these is differential privacy: a privacy notion that has attracted significant attention since its introduction in 2006 by Dwork, McSherry, Nissim, and Smith \cite{DMNS06}.
A computation on a collection of data is differentially private if the outcome is essentially statistically independent of any individual datum.

Work leading to differential privacy demonstrated fundamental tradeoffs between privacy and utility when computing statistics on sensitive information.
In 2003, Dinur and Nissim considered executing noisy statistical queries on a database $\db$ of $n$ entries~\cite{DiNi}. They showed that if the noise magnitude is $o(\sqrt{n})$, then there exists a simple linear program reconstructing all but a small fraction of $\db$, a result that was further generalized and strengthened in~\cite{DMT07,DY08}.
These reconstruction attacks provided useful guidance in the theoretical development of a rigorous approach to privacy, and in particular differential privacy.

\ifsubmission
\else
\paragraph{Reconstruction in practice.}
\fi In this work, we apply a linear reconstruction attack on a statistical query system in the wild.
To the best of our knowledge, this is the first time that such an attack has been successfully applied to reconstruct data from a commercially-available statistical database system specifically designed to protect the privacy of the underlying data.

The attack was performed on the production system called Diffix~\cite{Diffix} using a real dataset deployed in test environment in the course of a bug bounty program by Aircloak. The goal of Diffix is to allow data analysts to perform an unlimited number of statistical queries on a sensitive database while protecting the underlying data and while introducing only minimal error. It is being advertised as an off-the-shelf, GDPR-compliant privacy solution, and the company reports that ``CNIL, the French national data protection authority, has already evaluated Diffix against the GDPR anonymity criteria, and have stated that Aircloak delivers GDPR-level anonymity'' \cite{AIR_anonymisation}.
\medskip

As we show, by answering unlimited, highly accurate statistical queries Diffix is vulnerable to linear reconstruction attacks.

\section{Diffix}
Diffix is a system that sits between a data analyst and a dataset. The data analyst issues counting queries using a restricted subset of SQL. For example,
  \begin{center}
  \begin{minipage}{.8\linewidth}
    \texttt{SELECT count(*) FROM loans \\
    WHERE status = `C' AND client-id BETWEEN 2000 and 3000}
  \end{minipage}
  \end{center}
Diffix executes a related query on the underlying dataset and computes the answer to the query along with some additional random error. The noisy answer is returned to the analyst.

A primary focus of Diffix's design is noise generation, which is a function of both the text of the query and the subset of the data included in that query.
The noise is sampled from a zero-mean normal distribution and, depending on the data, rounded to the nearest integer.
The standard deviation of the noise depends on the complexity of the query; each additional condition in the query introduces an additional layer noise of standard deviation 1.

In addition to adding Gaussian noise, Diffix employs a number of heuristic techniques to promote privacy.
To protect against an attacker who may try to average noise out by issuing many logically equivalent but syntactically distinct queries, Diffix restricts the use of certain SQL operators, especially math operators.
Other techniques include suppressing small counts, modify extreme values, and disallowing many SQL operators (including OR).

In order to accelerate development and testing of our linear reconstruction attack, we simulated  Diffix's noise addition and small-count suppression in MATLAB. The results in Section~\ref{sec:attacking-aircloak} use real query responses from Diffix, while those in Section~\ref{sec:experiments} use the simulation.

\ifsubmission
\else
\paragraph{The Challenge.}
\fi
From December 2017 to May 2018, Aircloak ran ``the first bounty program for anonymized data re-identification,'' offering prizes of up to \$5,000 for successful attacks \cite{AIR_challenge}. The company granted researchers access to five datasets through Diffix, along with documentation of the design and implementation of Diffix and complete versions of the datasets for analysis.
Researchers were allowed to use auxiliary information gleaned directly from the datasets in order to carry out their attacks.
We commend Aircloak for making Diffix available to privacy researchers and for their support throughout this work.

Aircloak measured the success of an attack using an effectiveness parameter $\alpha$ and a confidence improvement parameter $\kappa$. They have verified our attack to achieve the best possible parameters. In a recent blog post, Aircloak reported that ``Only two attack teams formulated successful attacks. \ldots~Fixes for both attacks have been implemented'' \cite{AIR_fix}
At the time of writing, we have not examined the new restrictions on the query language introduced to by Aircloak to counter these attacks.

\section{Implementing the Linear Reconstruction Attack}
\label{sec:attacking-aircloak}

The attack targets a dataset $\db$ of size $n$ database entries indexed by a set of unique identifiers $\ID$. Each entry has an associated value of a Boolean target attribute, $\att_\id$.
Each query $\q \subseteq [n]$ specifies a subset of entries, and the response $\ans_\q = \q(\db) + \err_\q$ is the sum of true value $\q(\db) = \sum_{\id \in \q} \att_\id$ and an error term $\err_{\q}$.
The errors are sampled from a zero-mean Gaussian distribution of standard deviation $\sigma$, then rounded to the nearest integer.
Each query $\q$ is a uniformly random subset of $[n]$. The set of all queries is denoted $Q$ and is of size $m$.

We implemented a linear reconstruction attack following the approach of \cite{DMT07} to find a candidate database $\db'$ minimizing the total error.
\cite{DMT07} was designed for a setting when some errors may be very significant, but typical errors are small. In contrast, the linear program of \cite{DiNi} is suitable when there is a bound on the maximum error magnitude.
Although we use the linear program of \cite{DMT07}, we deviate by using subset queries.
That work analyzes a number of other types of queries, including $\pm 1$ queries of the form $\q^{\pm}(\db) = \sum_{\id \in \q} \att_\id - \sum_{\id \not\in \q} \att_\id$ for subsets $\q \subseteq [n]$.
While these queries can be implemented using subset queries,\footnote{$\q^{\pm}(\db)  = \q(\db) - \q^c(\db)$} the standard deviation of the resulting noise would be larger.
In contrast, subset queries were directly implementable in Diffix with less noise and proved effective.
Section~\ref{sec:experiments} reports on additional experiments testing the accuracy of these three contrasting approaches in the face of Gaussian noise.

We solve the following linear program over $n+m$ variables $\db' = (\att'_\id)_{\id\in\ID}$ and $(\err'_\q)_{\q\in Q}$:

\begin{center}
\begin{minipage}{.6\linewidth}
\begin{algorithm}[H]
  \SetKwInput{KwVars}{variables}
  \SetKwInput{KwMin}{minimize}
  \SetKwInput{KwSt}{subject to}
  \KwVars{$\db' = (\att'_\id)_{\id \in \ID}$ and $(\err'_\q)_{\q\in Q}$}
  \KwMin{$\sum_{\q \in Q} |\err'_\q|$}
  \KwSt{\hfill\\
\begin{center}
  \begin{tabular}{ll} $\forall q \in Q$, &$\err'_\q = \ans_\q - \q(\db')$\\
    $\forall \id\in\ID$, & $0 \le \att'_\id \le 1$
  \end{tabular}
\end{center}}
 \label{alg:dmt}
\end{algorithm}
\end{minipage}
\end{center}

\noindent There is a standard linearizing of the above nonlinear objective function by introducing $m$ additional variables.
To compute the final output, we round the real-valued $\att'_\id$ to the nearest value in $\zo$.

The results described in this section are from a reanalysis of data gathered during the Aircloak Challenge using the linear program described above. During the course of the actual challenge, we used a slightly modified linear program as described in Appendix~\ref{sec:actual-attack}.

\ifsubmission
\else
\paragraph{Querying ``random'' subsets.}
\fi
The main hurdle in implementing the attack was specifying queries for random subsets of the rows of the dataset.
Diffix determines the error magnitude per query depending on the description of the query. It increases the  noise magnitude for each additional condition in the query string. Random queries would require lengthy description and Diffix would hence introduce large noise that would reduce the reconstruction accuracy. We needed to find a way to specify a random---or ``random'' enough---subset of the data using as few conditions as possible.

Our approach, ad hoc yet ultimately effective, was to use the unique user identifier $\id$ as the source of ``randomness.'' For each "random" query we used a predicate $p_\q$ and let $q=\{\id: p_\q(\id)=1\}$.  Concretely, each query was specified by a prime $p$, an offset $j$, an exponent $e\in\{0.5,0.6,\dots,1.9\}\setminus\{1\}$, and a modulus $m\in\{2,5\}$. Row number $\id$ was included in the query $q = (p,j,e,m)$ if the $j$th digit in the decimal representation of $(p\cdot\id)^e$ was congruent to $0\mod m$.
For example, the following query corresponds to $p=2$, $j=2$, $e = 0.7$ and $m = 5$.
\begin{center}
\begin{minipage}{.8\linewidth}
  \texttt{SELECT count(clientId) FROM loans \\
    WHERE
    floor(100 * ((clientId * 2)$^\wedge$0.7) + 0.5) \\
    \textcolor{white}{hide}= floor(100 * ((clientId * 2)$^\wedge$0.7))}
\end{minipage}
\end{center}
The exact form of the query depended on the various syntactic restrictions included in Diffix.
By modifying the ranges of $p$ and $j$, we were able to tune the total number of queries. We restricted $p$ to the first 25 primes and $j\in [5]$, resulting in a total of 3500 queries.

\ifsubmission
\else
\paragraph{Results.}
\fi
Our target was the $\loans$ table in the $\banking$ dataset, consisting of real data of 827 loans from a bank in the Czech Republic.
The rows are indexed by the $\clientid$ attribute, a unique number between 2 and 13971.
Each row has an associated $\loanstatus$ attribute, a letter from `A' to `D'.

Our goal was to determine which loans had  $\loanstatus=\mbox{`C'}$, given only knowledge of the $\clientids$. In order to minimize the total number of queries, we restricted our attention to the subset of $\clientids$ in the range $[2000,3000]$, which contained 73 entries. Ultimately, our queries were of the form:
\begin{center}
\begin{minipage}{.8\linewidth}
  \texttt{SELECT count(clientId) FROM loans \\
    WHERE
    floor(100 * ((clientId * 2)$^\wedge$0.7) + 0.5) \\
    \textcolor{white}{hide}= floor(100 * ((clientId * 2)$^\wedge$0.7))\\
    AND clientId BETWEEN 2000 and 3000 \\
    AND loanStatus = `C'
    }
\end{minipage}
\end{center}
Diffix added error of standard deviation 4 to the output of these queries.
We applied the same attack on different ranges of $\clientids$ with 110, 130, and 142 entries (and in the last case, targeting the $\loanstatus$ value 'A'). In each case, we performed 3500 queries.

\bigskip

The linear program reconstructed the data for all four $\clientid$ ranges perfectly.

\section{Simulated Experiments}

\label{sec:experiments}

In addition to the above results using the actual Diffix system, we performed additional experiments using a simulation of a noisy statistical query mechanism. The simulated mechanism answers counting queries with zero-mean, normally-distributed noise with standard deviation $\sigma$ (and rounds to the nearest integer). It also suppresses low counts in the same way as the Diffix system, though that was only be relevant for the first experiment. All experiments described below were implemented in MATLAB on a personal laptop, and all linear programs were solved in less than 4 seconds.

\ifsubmission
\else
\paragraph{Removing auxiliary information.}
\fi
One drawback of our original attack on Diffix was the need for complete knowledge of the $\clientids$ as a prerequisite to performing the attack. Our first experiment sought to infer these $\clientids$. First, we identified a range of 100 possible $\clientids$ that had a large number of present $\clientids$ (relative to the other possible ranges). We want a large number of present $\clientids$ to minimize the effect of Diffix's low-count suppression. We settled on the range $[2500,2600]$ with 12 $\clientids$. While we identified this range using exact counts, we believe such a range could be found by querying Diffix itself.\footnote{{\em E.g.}, by issuing the query  \texttt{SELECT count(*) FROM loans WHERE clientid BETWEEN $a$ and $b$} to approximate the number of present $\clientids$ in the range $\{a,\ldots,b\}$.}

We simulated responses to 3500 queries of the following form:
\begin{center}
\begin{minipage}{.8\linewidth}
  \texttt{SELECT count(clientId) FROM loans \\
    WHERE
    floor(100 * ((clientId * 2)$^\wedge$0.7) + 0.5) \\
    \textcolor{white}{hide}= floor(100 * ((clientId * 2)$^\wedge$0.7))\\
    AND clientId BETWEEN 2500 and 2600 \\
    }
\end{minipage}
\end{center}
The \cite{DMT07} linear program was used to infer which $\clientids$ are present in the range. There was 1 false negative among the 12 present $\clientids$ and 0 false positives among the 88 absent $\clientids$.

\ifsubmission
\else
\paragraph{How accuracy varies with size, queries, and error.}
\fi
The accuracy of the linear reconstruction attack depends on the size of the dataset, the magnitude of the error, and the number of queries. When implementing our attack on Diffix, we used many more queries than seemed necessary for the level of noise used.
The next experiment illustrates how the accuracy of \cite{DMT07} varies with each of these parameters against a system using Gaussian noise to answer counting queries.

\begin{figure}[h]
    \centering
    \begin{subfigure}[b]{0.47\textwidth}
        \includegraphics[width=\textwidth]{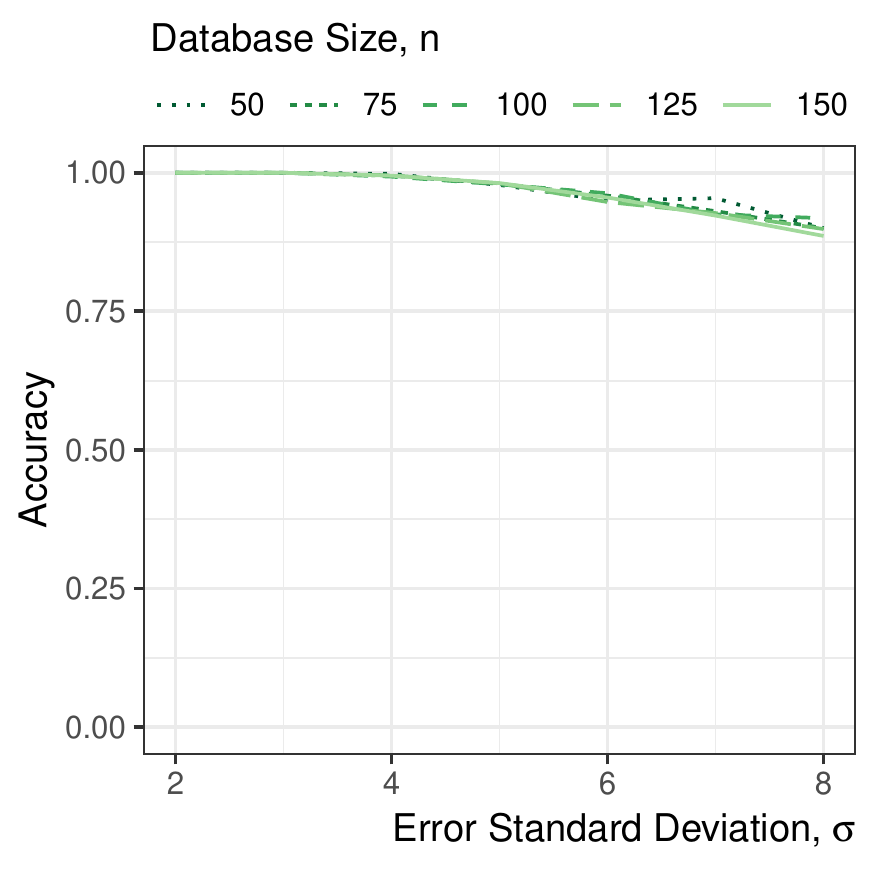}
        \caption{Noise standard deviation varied from 1 to 20, in increments of 1, with 2550 queries. For $n=100$, the mean accuracy falls below 0.99 at $\sigma = 5$ and below 0.95 at $\sigma = 7$.}
        \label{fig:errors-dmt}
    \end{subfigure}
    \quad
    \begin{subfigure}[b]{0.47\textwidth}
        \includegraphics[width=\textwidth]{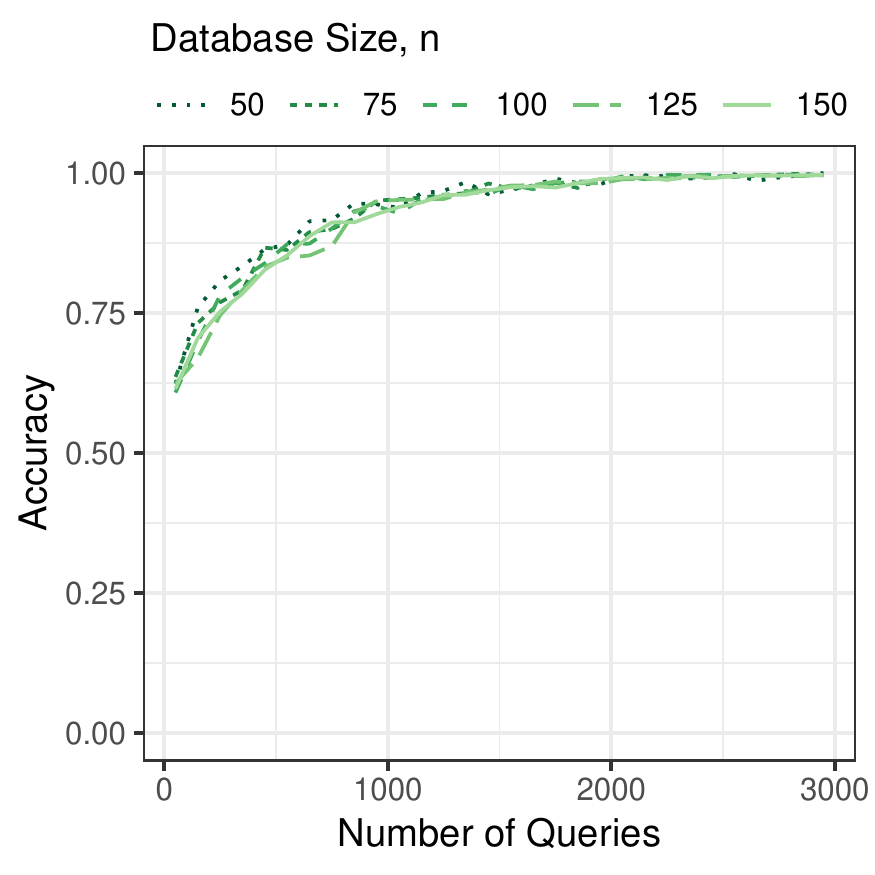}
        \caption{The number of queries varied from 50 to 2950, in increments of 100, with noise magnitude $\sigma = 4$. For $n=100$, the mean accuracy surpasses 0.95 at 1150 queries and surpasses 0.99 at 2050 queries.}
        \label{fig:queries-dmt}
    \end{subfigure}
    \caption{Reconstruction accuracy as a function of the (\ref{fig:errors-dmt}) noise magnitude and (\ref{fig:queries-dmt}) number of queries, for various database sizes. The data is averaged over 10 trials of the \cite{DMT07} linear program using subset queries.}\label{fig:dmt-experiments}
\end{figure}

The results are summarized in Figure~\ref{fig:dmt-experiments}.
The plots display the average accuracy over 10 simulated runs of our \cite{DMT07} reconstruction algorithm as the error magnitude, database size, and number of queries were varied. Each run resampled the Gaussian noise while the underlying dataset remained fixed. It is interesting to observe that the size of the dataset does not seem to significantly affect the effectiveness of reconstruction.

\begin{figure}[h]
    \centering
    \begin{subfigure}[b]{0.47\textwidth}
        \includegraphics[width=\textwidth]{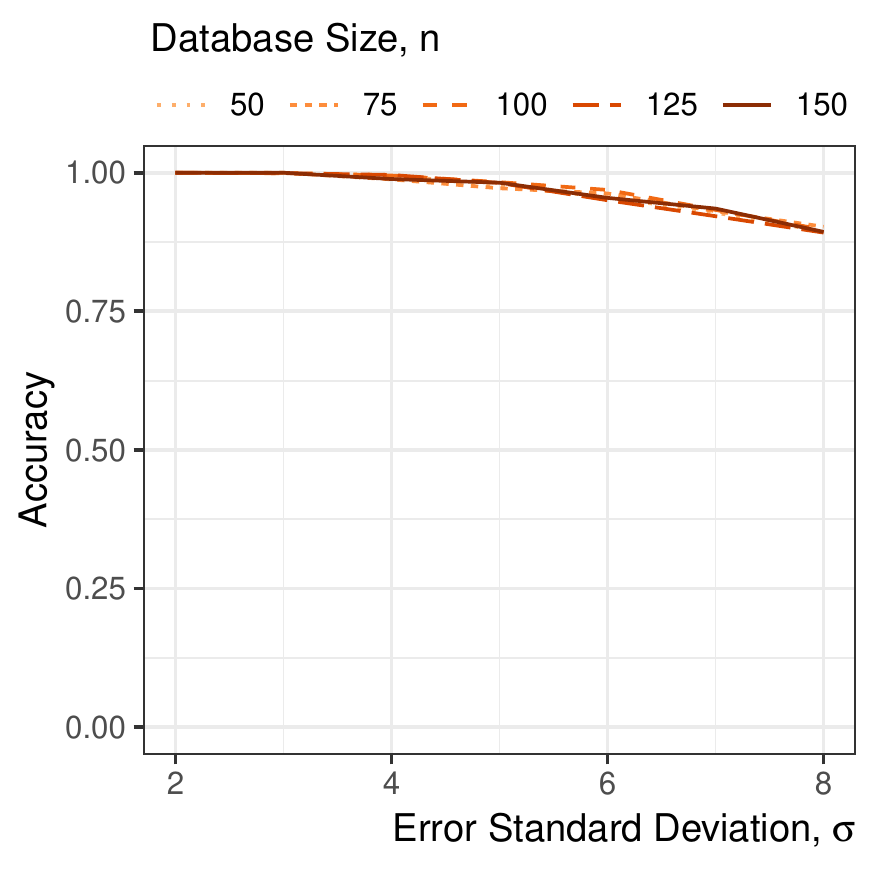}
        \caption{Noise standard deviation varied from 1 to 20, in increments of 1, with 2550 queries. For $n=100$, the mean accuracy falls below 0.99 at $\sigma = 5$ and below 0.95 at $\sigma = 8$.}
        \label{fig:errors-dmt-signed}
    \end{subfigure}
    \quad
    \begin{subfigure}[b]{0.47\textwidth}
        \includegraphics[width=\textwidth]{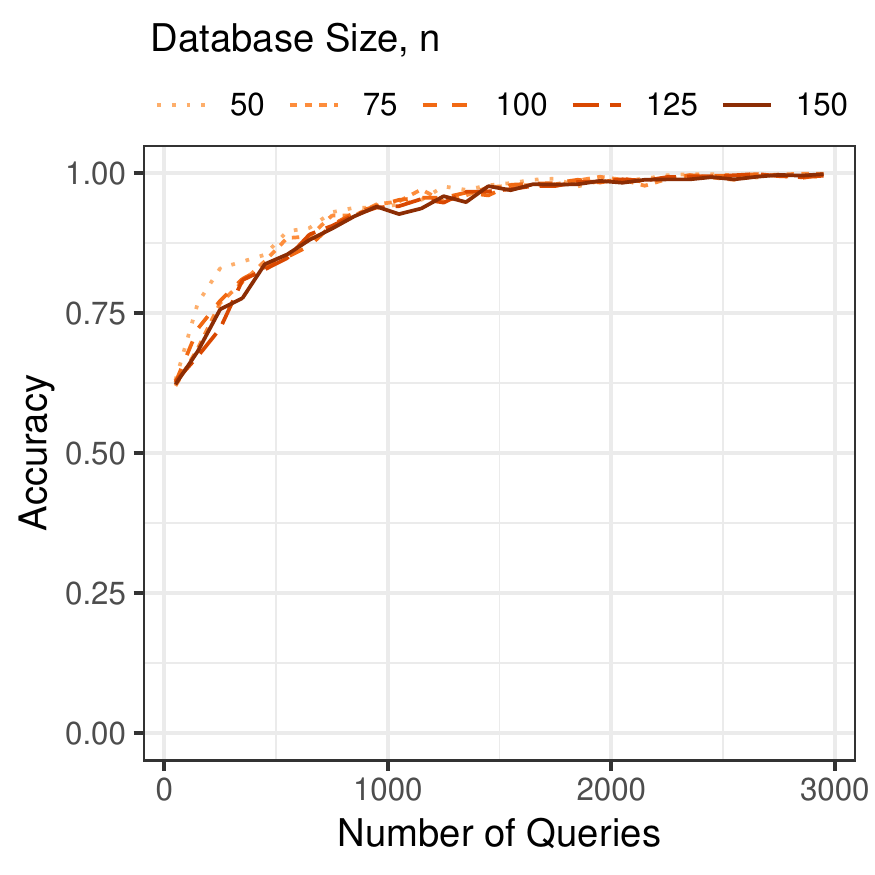}
        \caption{The number of queries varied from 50 to 2950, in increments of 100, with noise magnitude $\sigma = 4$. For $n=100$, the mean accuracy surpasses 0.95 at 1050 queries and surpasses 0.99 at 2050 queries.}
        \label{fig:queries-dmt-signed}
    \end{subfigure}
    \caption{Reconstruction accuracy as a function of the (\ref{fig:errors-dmt-signed}) noise magnitude and (\ref{fig:queries-dmt-signed}) number of queries, for various database sizes. The data is averaged over 10 trials of the \cite{DMT07} linear program using $\pm 1$ queries.}\label{fig:dmt-signed-experiments}
\end{figure}

As described in Section~\ref{sec:attacking-aircloak}, the analysis in \cite{DMT07} applies to $\pm 1$ queries but not to subset queries.
To compare the effectiveness of these two query types, we ran the same simulations using $\pm 1$ queries. The results are summarized in Figure~\ref{fig:dmt-signed-experiments}. The plots are nearly indistinguishable from the corresponding plots in Figure~\ref{fig:dmt-experiments}.

\ifsubmission
\else
\paragraph{Comparing \cite{DiNi} and \cite{DMT07}.}
\fi
The original linear reconstruction attack for noisy counting queries comes from \cite{DiNi}.
In contrast to \cite{DMT07}, \cite{DiNi} makes the additional assumption that each error $\err_\q$ is bounded by a maximum error $\Err$.
In our experiments, we write $\Err = B\sigma$, where $B$ is the \emph{error bound multiplier} and $\sigma$ is the standard deviation of the Gaussian errors.
The \cite{DiNi} linear program reflects the bounded-error assumption with an additional constraint and uses a trivial objective function.

\begin{center}
\begin{minipage}{.6\linewidth}
\begin{algorithm}[H]
  \SetKwInput{KwVars}{variables}
  \SetKwInput{KwMin}{minimize}
  \SetKwInput{KwSt}{subject to}
  \KwVars{$\db' = (\att'_\id)_{\id \in \ID}$ and $(\err'_\q)_{\q\in Q}$}
  \KwMin{0}
  \KwSt{\hfill\\
\begin{center}
  \begin{tabular}{ll} $\forall q \in Q$, &$\err'_\q = \ans_\q - \q(\db')$\\
    & $\err'_\q \le B\sigma$ \\
    $\forall \id\in\ID$, & $0 \le \att'_\id \le 1$
  \end{tabular}
\end{center}}
\end{algorithm}
\end{minipage}
\end{center}

\begin{figure}[h]
    \centering
    \begin{subfigure}[b]{0.47\textwidth}
        \includegraphics[width=\textwidth]{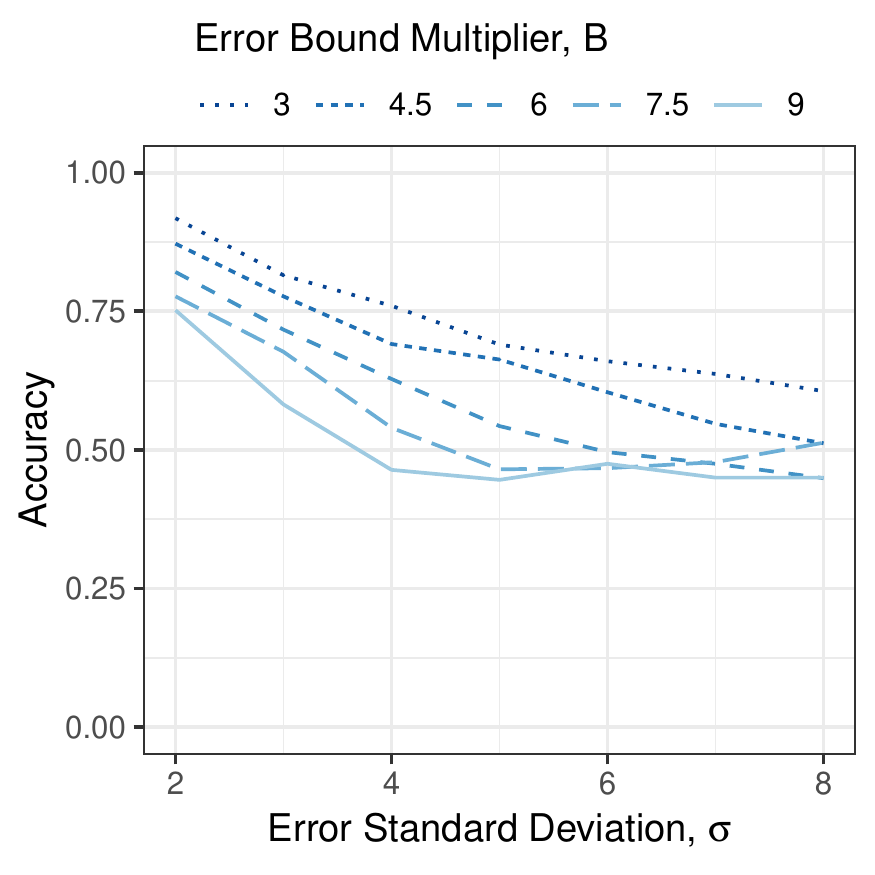}
        \caption{Noise standard deviation varied from 1 to 20, in increments of 1, with 2550 queries and dataset size $n = 100$.}
        \label{fig:errors-dini}
    \end{subfigure}
    \quad
    \begin{subfigure}[b]{0.47\textwidth}
        \includegraphics[width=\textwidth]{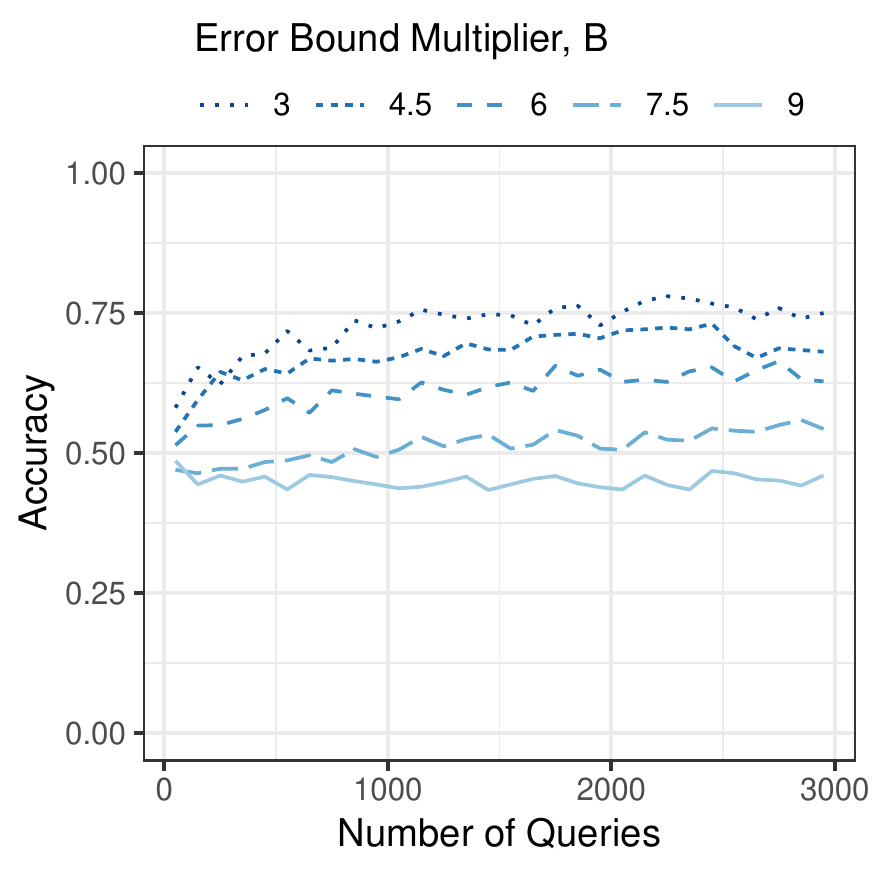}
        \caption{The number of queries varied from 50 to 2950, in increments of 100, with noise magnitude $\sigma = 4$ and dataset size $n = 100$.}
        \label{fig:queries-dini}
    \end{subfigure}
    \caption{Accuracy as a function of the (\ref{fig:errors-dini}) noise magnitude and (\ref{fig:queries-dini}) number of queries, for various values of the DiNi multiplier $B$. The data is averaged over 10 trials using a dataset of size $n = 100$.}\label{fig:dini-experiments}
\end{figure}

Our final experiment illustrates how the accuracy of the above $\cite{DiNi}$-based linear program varies as a function of the error magnitude, number of queries, and the error bound multiplier.
The results are summarized in Figure~\ref{fig:dini-experiments}.
The plots display the average accuracy over 10 simulated runs of our \cite{DiNi} reconstruction algorithm as the parameters were varied. Each run resampled the Gaussian noise while the underlying dataset remained fixed.

Observe that as the error bound multiplier $B$ increases, the accuracy of reconstruction degrades.
Because the linear program terminates once any feasible points is found, it is not surprising that expanding the set of feasible points by increasing $B$.

Note however that the pattern extends to $B = 3$. One would expect a few queries (in expectation about 4.6 queries per 2550 for $\sigma = 4$) to have rounded error greater than $3\sigma$. Nevertheless, in each of the 240 trials run with $B = 3$ and at least 2550 queries, a feasible solution was found.
In contrast, for $B=2.5$ and $\sigma = 4$ half of all executions with 1850 queries were infeasible (dropping to $\ge90\%$ infeasible at 2250 or more queries).

\bibliographystyle{alpha}			
\bibliography{bib}

\appendix

\section{Additional Information on the Aircloak Challenge Attack}
\label{sec:actual-attack}

The results described in Section~\ref{sec:attacking-aircloak} are from a reanalysis of data gathered during the Aircloak
Challenge.
During the course of the Aircloak Challenge, we used a modified version of the \cite{DMT07} linear program. For transparency, this section describes the modified linear program and its effectiveness.

The only difference between the linear program originally used and the one described in Section~\ref{sec:attacking-aircloak} is the addition of constraints upper bounding the magnitude of any error term.
\begin{center}
\begin{minipage}{.6\linewidth}
\begin{algorithm}[H]
  \SetKwInput{KwVars}{variables}
  \SetKwInput{KwMin}{minimize}
  \SetKwInput{KwSt}{subject to}
  \KwVars{$\db' = (\att'_\id)_{\id \in \ID}$ and $(\err'_\q)_{\q\in Q}$}
  \KwMin{$\sum_{\q \in Q} |\err'_\q|$}
  \KwSt{\hfill\\
\begin{center}
  \begin{tabular}{ll} $\forall q \in Q$, &$\err'_\q = \ans_\q - \q(\db')$\\
    & $\err'_\q \le 5\sigma$ \\
    $\forall \id\in\ID$, & $0 \le \att'_\id \le 1$
  \end{tabular}
\end{center}}
\end{algorithm}
\end{minipage}
\end{center}
\noindent where $\sigma = 4$ is the standard deviation of the true error distribution. Note that if the true errors were distributed according to $N(0,\sigma)$ and rounded to the nearest integer, an error of magnitude greater than $5\sigma$ would be expected once in every 1.7 million queries.

We first implemented the linear reconstruction solver using data from the $\clientid$ range $[2000,3000]$, for which it achieved perfect reconstruction.
Together with researchers at the Max Planck Institute for Software Systems, we verified the attack on three additional ranges of $\clientids$ containing 110, 130, and 142 $\clientids$. The results are summarized in Table~\ref{table:results}.
In two of the three ranges, the attack again inferred whether each $\loanstatus$ was `C' with high accuracy (1 and .9538).
We were surprised, therefore when the final validation (this time targeting $\loanstatus$ `A' rather than `C') achieved accuracy of only 75.4\%. Our confusion compounded when the accuracy degraded after increasing the number of queries, suggesting that we were not accounting for some source of error.

After further investigation, we realized that performing the queries for $\clientids$ in $[10000,12000]$ required more numerical precision than seemed to be supported by Diffix. The larger $\clientid$ values in this range and the larger constants required for additional queries introduced errors that had not affected our earlier tests. Ultimately, high accuracy was recovered by ignoring the results from queries with larger values of $e$ which seemed to require greater (but making no other changes to the linear program solver).

\begin{table}[h]
    \begin{center}
    \begin{tabular}{| c | c | c | c | c |}
    \hline
    $\clientids$ Range & Number of entries ($n$) & Number of queries & Target status & Accuracy \\ \hline \hline
    2000-3000 & 73 & 3500 & `C' & 1 \\ \hline
    3000-5000 & 110 & 3500 & `C' & 1 \\ \hline
    5000-7000 & 130 & 3500 & `C' & .9538 \\ \hline
    10000-12000 & 142 & 3500 & `A' & .7535 \\ \hline
    10000-12000 & 142 & 2000, $e \le 1.4$ & `A' & 1 \\ \hline
    10000-12000 & 142 & 1000, $e \le 0.8$ & `A' & .9930 \\ \hline
    \end{tabular}
    \caption{ \label{table:results} Summary of reconstruction tests performed against Diffix using the modified \cite{DMT07} linear program. The queries in the final two resulted from restricting the exponent $e$ to the indicated range.}
    \end{center}
\end{table}
\end{document}